\documentclass[review]{elsarticle}

\usepackage{lineno,hyperref}
\usepackage{amsmath}
\usepackage{amsthm}
\usepackage{amsfonts}
\usepackage{graphicx}
\usepackage{subfigure}
\usepackage{parskip}
\usepackage{siunitx}
\usepackage{arydshln}

\journal{Journal of Computational Physics}

\begin{document}

\begin{frontmatter}

\title{Explicit high-order symplectic integrators of coupled Schrödinger equations for pump-probe systems}
\author[USTC]{Xiaobao Jia}
\author[USTC]{Qing Jia\corref{mycorrespondingauthor}}
\author[USTC]{Jianyuan Xiao}
\author[USTC,IFSA]{Jian Zheng\corref{mycorrespondingauthor}}
\cortext[mycorrespondingauthor]{Corresponding author}
\ead{qjia@ustc.edu.cn, jzheng@ustc.edu.cn}
\address[USTC]{Department of Plasma Physics and Fusion Engineering and CAS Key Laboratory of Geospace Environment, University of Science and Technology of China, Hefei, Anhui 230026, People's Republic of China}
\address[IFSA]{Collaborative Innovation Center of IFSA, Shanghai Jiao Tong University, Shanghai 200240, People's Republic of China}
\begin{abstract}
Two-beam coupling within the field of nonlinear optics, which transfers energy from one light beam to the other under certain conditions, has received considerable attention in inertial confinement fusion (ICF) and plasma optics.
To evaluate the coupling dynamics precisely, we modeled this process with full-wave coupled Schrödinger equations (CSEs) and a nonlinear refractive index. We found that the CSEs constituted a Hamiltonian system and proposed an arbitrary higher-order explicit symplectic algorithm to solve the CSEs numerically.
The numerical results given by the developed BEAM code showed a good agreement with those from particle-in-cell simulations, which demonstrated the validity of the model and algorithm.
The model and numerical algorithm presented in this work can be extended to more nonlinear optical interactions described by coupled-wave equations.
\end{abstract}

\end{frontmatter}
\section{Introduction}
Nonlinear optical interaction is a dominant feature that distinguishes nonlinear optics from linear optics.
In recent two decades, there has been increasing interest in manipulating one light with another, which is based on the properties of nonlinear media, e.g., photorefractive effect, electrostrictive effect, acousto-optic effect, magneto-optic effect, electro-optic effect, etc\cite{boyd_nonlinear_2020}. As one of these nonlinear media, plasma is an important concern of nonlinear optical interactions, because it has an intrinsically nonlinear response to an intense applied optical field and there is no damage threshold. Nonlinear optical interactions in plasma offer new opportunities for optical components applicable to high intensity and high fluences, such as short pulse amplifier\cite{malkin_fast_1999}, plasma-based beam combiner\cite{kirkwood_plasma-based_2018}, and polarization control\cite{michel_dynamic_2014,michel_polarization-dependent_2020}.

The two-beam coupling (or two-wave mixing) is the consequence of the photorefractive effect\cite{boyd_nonlinear_2020} in a nonlinear medium. This physical process in plasma is as follows: firstly, the pondermotive force of two crossing lasers (pump and probe beams) generates a density modulation with the interference pattern structure. Then, the density modulation evolves into a beating plasma wave, which scatters each beam in the direction of the other, thus affecting the phase of each laser beam. The phase shift contribution of the pump (probe) beam to the probe (pump) beam, related to the relative phase between two beams\cite {wahlstrand_effect_2013}, is usually expressed as a complex coupling coefficient. Due to the phase shift caused by scattering, the probe (pump) beam undergoes a phase modulation corresponding to the imaginary part of the complex coefficient or an amplitude modulation corresponding to the real part.

The two-beam coupling in plasmas is of great significance both in inertial confinement fusion (ICF) and in plasma optics. In ICF, crossed-beam energy transfer (CBET)\cite{kruer_energy_1996} have a deleterious effect in direct-drive ICF while it is utilized to adjust the laser energy between the inner and the outer cones in indirect-drive \cite{myatt_multiple-beam_2014}. The CBET problem was used to be modeled by the ray-based methods, which treat the electromagnetic wave as a bundle of rays of plane wave and might lose the phase information of practical lasers. The geometric optics approximation also fails in the vicinity of caustics\cite{follett_full-wave_2017,follett_ray-based_2018}. In plasma optics, the plasma-based optical components demand appropriate physical models and numerical tools. In this work, we describe the two-beam coupling by a coupled-wave model avoiding the failure of geometric optics and is applicable for plasma optics.

In this paper, we model the two-beam coupling with the coupled Schrödinger equations (CSEs) deduced from the coupled-wave equations in a slowly varying envelope approximation. We find that the CSEs constitute a Hamiltonian system, which preserves its intrinsic properties after the application of symplectic integrators\cite{feng_symplectic_1987}.
Then, we propose an arbitrary high-order explicit symplectic algorithm resorting to the Hamiltonian splitting method and develop a matching code BEAM of second-order accuracy. We apply the BEAM code to demonstrate the physics of the Crossed-Beam Energy Transfer (CBET) in ICF and Brillouin amplification, which shows good agreement with particle-in-cell (PIC) simulation but with a much lower computational cost. The significance and originality of this work are that it provides an efficient conservative and fast numerical algorithm or solution idea for simulating long-time and large-scale nonlinear optical interactions described by coupled-wave equations since they have symmetric coupling items in general.

This paper is organized as follows. In Section 2, we derive the coupled Schrödinger equations (CSEs) as the theoretical model for two-beam coupling in plasma and other nonlinear media. In Section 3, we deduce the Hamiltonian of coupled Schrödinger equations and propose the arbitrary high-order explicit symplectic algorithm resorting to the Hamiltonian splitting method. In Section 4, to demonstrate the validity of the model and numerical algorithm, numerical examples obtained from BEAM1D/2D are presented, and the results are compared with kinetic simulations. Finally, we summarize this work in Section 5.

\section{Theoretical model}
\subsection{Coupled wave equations}
We examine two electromagnetic (EM) waves that interact in a plasma with a background density $n_0$. A nonuniform density modulation will be generated due to the pondermotive force of the superposition of light fields. For quasi-monochromatic waves with mean frequencies $\omega_1,\,\omega_2$, the spectral bandwidth satisfies $\Delta\omega_{1, 2}\ll\omega_{1, 2}$, and the electrical fields can be expressed as
\begin{equation}
	\begin{aligned}
\mathbf{E_1}(\mathbf{r},t)&={E}_{1,s}(\mathbf{r},t)e^{-i\omega_1 t}\mathbf{\hat{e}_{s}}+{E}_{1,p}(\mathbf{r},t)e^{-i\omega_1 t}\mathbf{\hat{e}_{p}}+c.c.,\\
\mathbf{E_2}(\mathbf{r},t)&={E}_{2,s}(\mathbf{r},t)e^{-i\omega_2 t}\mathbf{\hat{e}_{s}}+{E}_{2,p}(\mathbf{r},t)e^{-i\omega_2 t}\mathbf{\hat{e}_{p}}+c.c.,
	\end{aligned}\label{eq1}
\end{equation}
where $E_{1,s}$, $E_{1,p}$, $E_{2,s}$ and ${E_{2,p}}$ are the projections of two complex enveloped electric field in s and p directions, $\mathbf{\hat{e}_{s}, \ \hat{e}_{p}}$ are two orthogonal unit vectors and the $c.c.$ stands for the complex conjugate of the complex variables.
Here we also introduce the complex variable $n_1$ caused by the interference pattern of these two EM waves,
\begin{equation}
 n_1=n_{1, R}+in_{1, I}=|n_1|e^{i(\phi_3-\omega_3 t+\Delta \phi)},\ \delta n=Re[n_1]=\frac{n_1+n_1^{*}}{2}\ (|\delta n|\ll n_0),\label{eq2}
 \end{equation}
where the difference frequency $\omega_3=\omega_1-\omega_2$ satisfies $\omega_3\ll \omega_{1, 2}$ and $\delta n$ is the density perturbation relative to the ambient plasma density $n_0$ hence the total density $n=n_0+\delta n$. Due to the intrinsic relaxation time of plasmas, a response delay of plasma inducing the dephasing term $\Delta \phi$ exists, and therefore the distribution of the density distribution does not consist with the laser intensity distribution in real-time.

After the definition of $n_1$ and enveloped electric fields, we begin with the wave equation of electric field in a plasma\cite{myatt_lpse_2019},
\begin{equation}
\nabla(\nabla\cdot \mathbf{E})-\nabla^2\mathbf{E}+\frac{1}{c^2}\frac{\partial^2\mathbf{E}}{\partial t^2}=-\frac{4\pi}{c^2}\frac{\partial \mathbf{J}}{\partial t},\label{eq4}
\end{equation}
where the induced current $\mathbf{J}$ is given by $\mathbf{J=\sigma_{1,2} I:E}$ and $\sigma_{1,2}= \frac{ie^2n}{m_e\omega_{1,2}}$ is the electrical conductivity.
Since the electrostatic force is not the leading role in driving the plasma density perturbation, and the spatial scale is much larger than the Debye length, the quasi-neutrality $(n_e\approx Zn_i,\ \rho\approx 0)$ condition is satisfied.
For simplicity, we consider the s-polarized EM waves, i.e., $E_{1,s}=E_1$ and $E_{2,s}=E_2$, and thus the first item on the left-hand side of Eq.\eqref{eq4} will vanish. Since the two lasers have the same polarization direction, the vector wave equation reduces into a scalar equation, which means that the applications our method is valid within the regime of scalar theory of optics. 

Replacing ${J}$ with $E$ by the conductivity
\begin{equation}
J=\frac{ie^2}{m_e}(n_0+\frac{n_1+n_1^{*}}{2})\left(\frac{E_1e^{-i\omega_1t}}{\omega_{1}}+\frac{E_2e^{-i\omega_2t}}{\omega_2}\right),\label{eq5}
\end{equation}
the two-beam coupling is accounted self-consistently in the nonlinear polarization. As is mentioned above, the nonlinear density modulation $|\delta n|$ is caused by the pondermotive force, so it implies that $|\delta n|\propto |E^2|$, which shows that the third-order nonlinear optical interactions are included in terms containing $n_1$ and $n_1^{*}$.

After substituting the total electrical field $E=E_1e^{-i\omega_1 t}+E_2e^{-i\omega_2 t}$ and \eqref{eq5} into \eqref{eq4} and applying the slowly varying envelope approximation $|\omega_{1, 2}{E}_{1, 2}|\gg |\frac{\partial {E}_{1, 2}}{\partial t}|$, we get
\begin{equation}
\begin{aligned}
\left( \nabla ^2-\frac{\omega _{pe}^{2}}{c^2} \right) \left( E_1e^{-i\omega _1t}+E_2e^{-i\omega _2t} \right) +i2\frac{\omega _1}{c^2}\frac{\partial E_1}{\partial t}e^{-i\omega _1t}+\frac{\omega _1^2}{c^2}E_1e^{-i\omega _1t}
\\
+i2\frac{\omega _2}{c^2}\frac{\partial E_2}{\partial t}e^{-i\omega _2t}+\frac{\omega _2^2}{c^2}E_2e^{-i\omega 2t}=\frac{\omega _{pe}^{2}}{c^2}\frac{n_1+n_{1}^{*}}{2}\left( E_1e^{-i\omega _1t}+E_2e^{-i\omega _2t} \right) ,\label{eq6}
\end{aligned}
\end{equation}
where $\omega_{pe}=\frac{4\pi e^2 n_0}{m_e}$ is the plasma frequency.
Equation \eqref{eq6} consists of multiple phase components which satisfy the phase matching condition\cite{boyd_nonlinear_2020}, in our model, if we let the phase of the complex enveloped fields $E_1$ and $E_2$ are $\phi_1$ and $\phi_2$ respectively, and then the phase condition is $\omega_1-\omega_2=\omega_3,\ \phi_1-\phi_2=\phi_3$. Extracting the subequations according to the phase $\phi_1$ and $\phi_2$, we may write
\begin{subequations}
	\begin{align}
\nabla ^2E_1+i2\frac{\omega _1}{c^2}\frac{\partial E_1}{\partial t}+\frac{\omega _1^2}{c^2}E_1-\frac{\omega _{pe}^{2}}{c^2}E_1&=\frac{\omega _{pe}^{2}}{c^2}\frac{n_1e^{i\omega _3t}}{2}E_2,\label{eq7a}
\\
\nabla ^2E_2+i2\frac{\omega _2}{c^2}\frac{\partial E_2}{\partial t}+\frac{\omega _2^2}{c^2}E_2-\frac{\omega _{pe}^{2}}{c^2}E_2&=\frac{\omega _{pe}^{2}}{c^2}\frac{n_{1}^{*}e^{-i\omega _3t}}{2}E_1.\label{eq7b}
	\end{align}
\end{subequations}
So far we have obtained equations describing the nonlinear optical interaction, it follows the normalization of Eq.\eqref{eq7a}-\eqref{eq7b} as 
\begin{equation}
t=\frac{\bar{t}}{\omega_1},\ x(y, z)=\frac{c\bar{x}(\bar{y},\bar{z})}{\omega_1},\ n_0 (n_1, n_1^{*})=n_c\bar{n}_0 (\bar{n}_1, \bar{n}_1^{*}),\ n_c=\frac{m_e\omega_1 }{4\pi e^2},\label{eq8}
\end{equation}
where $n_c$ is the critical density with respect to $\omega_1$. Since all the variables apart from the enveloped electric field $E_{1,2}$ have been normalized, we omit the bars which stand for the normalized quantities, and in this way, the subequations become
\begin{subequations}
	\begin{align}
{\nabla}^2E_1+i2\frac{\partial E_1}{\partial {t}}+\left( 1-{n}_0 \right) E_1&=\frac{{n}_1e^{i\omega _3t}}{2}E_2,\label{eq9a}
\\
{\nabla}^2E_2+i2\frac{\omega _2}{\omega _1}\frac{\partial E_2}{\partial {t}}+\left( \frac{\omega _2^2}{\omega _1^2}-{n}_0 \right) E_2&=\frac{{n}_{1}^{*}e^{-i\omega _3t}}{2}E_1.\label{eq9b}
	\end{align}
\end{subequations}
In most situations, since the frequency difference between two lasers is not too large, on the order of a few thousandths in the plasma, we can approximate $\frac{\omega_2}{\omega_1}$ and $\frac{\omega_2^2}{\omega_1^2}$ to 1. Then we can obtain the coupled Schrödinger equations as
\begin{subequations}
	\begin{align}
{\nabla}^2E_1+i2\frac{\partial E_1}{\partial {t}}+\left( 1-{n}_0 \right) E_1&=2KE_2,\label{eq10a}
\\
{\nabla}^2E_2+i2\frac{\partial E_2}{\partial {t}}+\left(1-{n}_0 \right) E_2&=2K^{*}E_1,\label{eq10b}
	\end{align}
\end{subequations}
where $K=\frac{\bar{n}_1e^{i\omega_3 t}}{4}$ represents the complex coupling coefficient.

\subsection{Nonlinear refractive index in a plasma}
When two-beam coupling happens in a plasma, the refractive index of plasma can be devided into linear and nonlinear parts:
\begin{equation}
N=\sqrt{(1-{n}_0)-{n}_1e^{i\Delta\phi}}=N_L+N_{NL},\ N_L=\sqrt{1-{n}_0}.\label{eq11}
\end{equation}
From Eq.\eqref{eq11}, we can approximate $N_{NL}=-\frac{{n}_1e^{i\omega_3 t}}{2N_L}=\frac{2K}{N_L}$. In a plasma, since the complex nonlinear refractive index ($N_{NL}\propto K$)is associated with the plasma density perturbation which forms a plasma wave, we can solve the complex density perturbation to obtain the coupling coefficients K.

Ion-acoustic wave is a low-frequency wave in an unmagnetized plasma and we can only consider the motion of ions in the condition of the quasi-neutrality. The equations governing the ion-acoustic wave are continuity equation, momentum equation, and the equation of state in terms of first-order small quantities, namely, the density and velocity perturbation $n_1$ and $\mathbf{u_1}$. This set of equations has been derived in previous literatures\cite{kruer_physics_1988,hittinger_simulating_2005,myatt_lpse_2019}, so we directly write the final form as
\begin{subequations} 
\begin{align}
\frac{\partial n_1}{\partial t}+\nabla\cdot\mathbf{u_1}+\mathbf{u_0}\cdot\nabla n_1&=0\label{eq12a}\\
\frac{\partial \mathbf{u_1}}{\partial t}+\mathbf{u_0\cdot \nabla u_1}+2\nu_a\mathbf{u_1}+c_s^2\nabla n_1&=-\frac{ZT_e}{m_i}\nabla[\bar{E}_1 \bar{E}_2^{*}e^{-i\omega_3 t}+0.5(|\bar{E}_1|^2+|\bar{E}_2|^2)]\label{eq12b}
\end{align}
\end{subequations}
where the bar denotes electric fields normalized by $\frac{\omega_1\sqrt{m_eT_e}}{e}$ ($m_i$, $T_e$ are the ion mass and the temperature of electrons), $\mathbf{u_0}$ is the velocity of the background plasma and $\nu_a$ stands for the damping rate of the ion-acoustic wave. In sec 2.1, we have given that the phase of $n_1$ is $\phi_3-\omega_3t$ which is consistent with the phase of pondermotive force. But in fact, the non-instantaneous response of plasma imposes a phase shift $\Delta\phi$ on $n_1$, and this is embodied in the time derivative terms of Eq.\eqref{eq12a}-\eqref{eq12b}.

The first term on the righthand side of Eq.\eqref{eq12b} is the driving force, the pondermotive force, of the dynamic plasma grating while the second term represents the self-focusing of laser beams in plasma only counts in the large spatial and temporal scales. Since the evolution time of the zero-order quantities $n_0$ and $\mathbf{u_0}$, i.e., the characteristic time scale of the hydrodynamics, is much longer than the period of the ion-acoustic waves, meanwhile, the spatial scale of the zero-order quantities is much larger than the wavelength of the ion-acoustic waves, we are allowed to take the zero-order quantities as constants in small spatial scale and short time scale. However, if the two-beam coupling happens on moderate or large spatial and temporal scales, it is necessary to take the evolution of $n_0$ and $\mathbf{u_0}$ and the effect of self-focusing into account since the feedback from the two-beam coupling to the hydrodynamic parameters becomes significant.

\subsection{Nonlinear refractive index in other nonlinear media}
In many nonlinear media, the refractive index depends on the intensity of the applied electric field. The steady-state intensity-dependent index can be described by\cite{boyd_nonlinear_2020,kur_nonlinear_2021}
\begin{equation}
N=N_L+N_{NL}=N_L+KI=N_L+\frac{N_Lc}{2\pi}|{E}_1+{E}_2|^2\label{eq13}
\end{equation}
where $I$ is the laser intensity averaging over fast oscillations. In this case, the nonlinear refractive index synchronizes with laser intensity distribution, which means no net energy transferring because of the instantaneous nonlinear response.

As for the transient state cases, where energy transfer will happen, the nonlinear refractive index becomes\cite{boyd_nonlinear_2020}
\begin{equation}
N_{NL}=\frac{N_Lc}{2\pi}\left({E}_1{E}_1^{*}+{E}_2{E}_2^{*}+\frac{{E}_1{E}_2^{*}}{1-i\omega_3\tau}+\frac{{E}_1^{*}{E}_2}{1+i\omega_3\tau}\right)\label{eq14}
\end{equation}
where $\tau$ is the response time of the nonlinear media.

\section{Numerical algorithm}
In this section, we prove that the coupled Schrödinger equations (CSEs) Eq.\eqref{eq10a}-\eqref{eq10b} constitute a Hamiltonian system and deduce its Hamiltonian. To preserve the structure of the Hamiltonian system, the explicit high-order symplectic integrators based on the Hamiltonian splitting method\cite{hairer_geometric_2006,yoshida_construction_1990} have been constructed. The coupling system governed by CSEs and equations of nonlinear refractive index is divided into six subsystems, which all have exact solutions. The high-order symplectic algorithm is just the composition of them. The Hamiltonian splitting guarantees iteration map of each subsystem is still Hamiltonian and of course the composition of these solutions preserves the symplectic structure associated with the original Hamiltonian system. Then, the numerical solution of the nonlinear refractive index is also presented.
\subsection{Hamiltionian of Coupled Schrödinger equations}
In Section 2.1, we have derived the CSEs and noticed that the coupling terms on the righthand side of Eq.\eqref{eq10a}-\eqref{eq10b} are symmetric. For simplicity, we use $\hat{H}$ to denote the spatial operator $-\frac 1 2 \nabla^2+\frac{1-{n}_0}{2}$ and drop the bar on the time derivative t, then the CSEs reduce to
\begin{subequations}
	\begin{align}
	i\frac{\partial E_1}{\partial t}&=\hat{H}E_1+KE_2\label{eq15a},\\
	i\frac{\partial E_2}{\partial t}&=\hat{H}E_2+K^{*}E_1\label{eq15b}.
	\end{align}
\end{subequations}
It is well-established that the Schrödinger equation without coupling terms is equivalent to a Hamiltonian canonical equation wherein the generalized coordinate and momentum are the real and imaginary parts of $E_{1, 2}$ respectively. To handle the symmetric coupling terms and find out the conservative quantity of the CSEs, we extract the real and imaginary parts of Eq.\eqref{eq15a}-\eqref{eq15b},
\begin{subequations}
	\begin{align}
	\frac{\partial E_{1, R}}{\partial t}&=\hat{H}E_{1, I}+K_{1, R}E_{2, I}+K_{1, I}E_{2, R}\label{eq16a},\\
	\frac{\partial E_{2, R}}{\partial t}&=\hat{H}E_{2, I}+K_{1, R}E_{1, I}-K_{1, I}E_{1, R}\label{eq16b},\\	
	\frac{\partial E_{1, I}}{\partial t}&=-\hat{H}E_{1, R}-K_{1, R}E_{2, R}+K_{1, I}E_{2, I}\label{eq16c},\\
	\frac{\partial E_{2, I}}{\partial t}&=-\hat{H}E_{2, R}-K_{1, R}E_{1, R}-K_{1, I}E_{1, I}\label{eq16d},
	\end{align}
\end{subequations}
where the additional subscript ``R'' and ``I'' represent the real part and imaginary part respectively.

If the elements of column vector $\mathbf{E}=(E_{1, R}, E_{2, R}, E_{1, I}, E_{2, I})^T$ are chosen as canonical variables in 4-dimensional phase space, the canonical system reads
\begin{equation}
\frac{\partial E_{m, R}}{\partial t}=\frac{\partial\mathcal{H}}{\partial{E_{m, I}}}, \frac{\partial E_{m, I}}{\partial t}=-\frac{\partial\mathcal{H}}{\partial{E_{m, R}}}, m=1, 2.\label{eq17}
\end{equation}
Comparing Eq.\eqref{eq16a}-\eqref{eq16d} with Eq.\eqref{eq17} allows us to write down $\mathcal{H}$
\begin{equation}
	\begin{aligned}
		\mathcal{H}=&E_{1, R}\hat{H}E_{1, R}+E_{1, I}\hat{H}E_{1, I}+E_{2, R}\hat{H}E_{2, R}+E_{1, I}\hat{H}E_{1, I}\\
		&+2K_{1, R}(E_{1, R}E_{2, R}+E_{1, I}E_{2, I})+2K_{1, I}(E_{1, I}E_{2, R}-E_{1, R}E_{2, I}),\label{eq18}
		\end{aligned}
\end{equation}
which is exactly the Hamiltonian of the coupled system. It follows that the CSEs form a Hamiltonian system which has a conservative quantity $\mathcal{H}$ because it does not depend explicitly on the time variable.

\subsection{Explicit high-order symplectic scheme based on Hamiltonian splitting}
Here we construct the high-order explicit symplectic algorithm by the Hamiltonian splitting method. Recalling the Hamiltonian in Eq.\eqref{eq18}, if we apply the Hamiltonian splitting $\mathcal{H=H_R+H_I+H_{A}+H_{B}+H_{C}+H_{D}}$, the original Hamiltonian system will be divided into six Hamiltonian subsystems which have exact solutions, and hence, an arbitrarily high-order symplectic algorithm is available by different composition methods.

With vector-valued canonical variable $\mathbf{E}=(E_{1, R}, E_{2, R}, E_{1, I}, E_{2, I})$, the six subsystems in the form of Poisson bracket are
\begin{subequations}
	\begin{align}
	\dot{\mathbf{E}}&=\{\mathbf{E}, \mathcal{H_R}\},\ \dot{\mathbf{E}}=\{\mathbf{E}, \mathcal{H_I}\},\ \dot{\mathbf{E}}=\{\mathbf{E}, \mathcal{H_A}\},\label{eq19a}\\
	\dot{\mathbf{E}}&=\{\mathbf{E}, \mathcal{H_B}\},\ \dot{\mathbf{E}}=\{\mathbf{E}, \mathcal{H_C}\},\ \dot{\mathbf{E}}=\{\mathbf{E}, \mathcal{H_D}\},\label{eq19b}
	\end{align}
\end{subequations} 
and 
\begin{subequations}
	\begin{align}
	\mathcal{H_R}&=E_{1, R}\hat{H}E_{1, R}+E_{2, R}\hat{H}E_{2, R},\label{eq20a}\\
	 \mathcal{H_I}&=E_{1, I}\hat{H}E_{1, I}+E_{2, I}\hat{H}E_{2, I},\label{eq20b}\\
	\mathcal{H_{A}}&=2K_{1, R}E_{1, R}E_{2, R},\ \mathcal{H_{B}}=2K_{1, R}E_{1, I}E_{2, I},\label{eq20c}\\
	\mathcal{H_{C}}&=-2K_{1, I}E_{1, R}E_{2, I},\ \mathcal{H_{D}}=2K_{1, I}E_{1, I}E_{2, R}.\label{eq20d}
	\end{align}
\end{subequations}
The first two Hamiltonian $\mathcal{H_R, H_I}$ represent the terms corresponding to the real and imaginary parts of electric fields, and $\mathcal{H_{A}, H_{B}, H_{C}, H_{D}}$ associate with terms related to the real and imaginary parts of coupling coefficients.

Then we start from the first subsystem $\dot{\mathbf{E}}=\{\mathbf{E}, \mathcal{H_R}\}$ and define the map from $\mathbf{E}(t)$ to $\mathbf{E}(t+\Delta t)$ as $\Theta_R(\Delta t)$. The transformation matrix $\mathcal{M_R}$ can be written as
\begin{equation}
	\mathbf{\dot{E}}=\mathcal{M_R}\mathbf{E},\ 
	\mathcal{M_R}=\left[
	\begin{array}{cc:cc}
	0 & 0 & 0 & 0\\
	0 & 0 & 0 & 0\\
	\hdashline
	-\hat{H} & 0 & 0 & 0\\
	0 & -\hat{H} & 0& 0
	\end{array}
\right],
\label{eq21}
\end{equation}
which shows that the real part ${E}_{m, R}$ does not depend on time and the imaginary part $E_{m, I}$ changes linearly with time. Therefore, the exact solution $\Theta_R(\Delta t)$ is
\begin{subequations}
	\begin{align}
	E_{1, R}(t+\Delta t)&=E_{1, R}(t),\label{eq22a}\\
	E_{2, R}(t+\Delta t)&=E_{2, R}(t),\label{eq22b}\\
	E_{1, I}(t+\Delta t)&=E_{1, I}(t)-\Delta t\hat{H}E_{1, R}(t),\label{eq22c}\\
	E_{2, I}(t+\Delta t)&=E_{2, I}(t)-\Delta t\hat{H}E_{2, R}(t).\label{eq22d}
	\end{align}
\end{subequations}
Similarly, we may write the $\mathcal{H_I}$ subsystem as
\begin{equation}
	\mathbf{\dot{E}}=\mathcal{M_I}\mathbf{E},\ 
	\mathcal{M_I}=\left[
	\begin{array}{cc:cc}
	0 & 0 & \hat{H} & 0\\
	0 & 0 & 0 & \hat{H}\\
	\hdashline
	0 & 0 & 0 & 0\\
	0 & 0 & 0 & 0
	\end{array}
\right].
\label{eq23}
\end{equation}
Accordingly, the exact solution $\Theta_I(\Delta t)$ reads
\begin{subequations}
	\begin{align}
	E_{1, I}(t+\Delta t)&=E_{1, I}(t),\label{eq24a}\\
	E_{2, I}(t+\Delta t)&=E_{2, I}(t),\label{eq24b}\\
	E_{1, R}(t+\Delta t)&=E_{1, R}(t)+\Delta t\hat{H}E_{1, I}(t),\label{eq24c}\\
	E_{2, R}(t+\Delta t)&=E_{2, R}(t)+\Delta t\hat{H}E_{2, I}(t).\label{eq24d}
	\end{align}
\end{subequations}
As for the $\mathcal{H_{A}}$ subsystem
\begin{equation}
	\mathbf{\dot{E}}=2K_{1, R}\mathcal{M_{A}}\mathbf{E},\ 
	\mathcal{M_{A}}=\left[
	\begin{array}{cc:cc}
	0 & 0 & 0 & 0\\
	0 & 0 &0 & 0\\
	\hdashline
	0 &-1 & 0 & 0\\
       -1 & 0 & 0 & 0
	\end{array}
\right],
\label{eq25}
\end{equation}
it has the exact solution $\Theta_A(\Delta t)$ as
\begin{subequations}
	\begin{align}
	E_{1, R}(t+\Delta t)&=E_{1, R}(t),\label{eq26a}\\
	E_{2, R}(t+\Delta t)&=E_{2, R}(t),\label{eq26b}\\
	E_{1, I}(t+\Delta t)&=E_{1, I}(t)-2\Delta t K_{1, R}E_{2, R}(t),\label{eq26c}\\
	E_{2, I}(t+\Delta t)&=E_{2, I}(t)-2\Delta t K_{1, R}E_{1, R}(t).\label{eq26d}
	\end{align}
\end{subequations}
The tramsformation matrix $\mathcal{M_B}$ and the exact solution $\Theta_B$ of $\mathcal{H_B}$ subsymtem appear as
\begin{equation}
	\mathbf{\dot{E}}=2K_{1, R}\mathcal{M_{B}}\mathbf{E},\ 
	\mathcal{M_{B}}=\left[
	\begin{array}{cc:cc}
	0 & 0 & 0 & 1\\
	0 & 0 &1 & 0\\
	\hdashline
	0 &0 & 0 & 0\\
       0 & 0 & 0 & 0
	\end{array}
\right],
\label{eq27}
\end{equation}
and 
\begin{subequations}
	\begin{align}
	E_{1, I}(t+\Delta t)&=E_{1, I}(t),\label{eq28a}\\
	E_{2, I}(t+\Delta t)&=E_{2, I}(t),\label{eq28b}\\
	E_{1, R}(t+\Delta t)&=E_{1, R}(t)+2\Delta t K_{1, R}E_{2, I}(t),\label{eq28c}\\
	E_{2, R}(t+\Delta t)&=E_{2, R}(t)+2\Delta t K_{1, R}E_{1, I}(t).\label{eq28d}
	\end{align}
\end{subequations}
Following a similar approach, we can write down the exact solution of $\mathcal{H_C}$ and $\mathcal{H_D}$ subsystem as following
\begin{subequations}
	\begin{align}
	\Theta_C(\Delta t):\quad E_{1, R}(t+\Delta t)&=E_{1, R}(t),\label{eq29a}\\
	E_{2, I}(t+\Delta t)&=E_{2, I}(t),\label{eq29b}\\
	E_{2, R}(t+\Delta t)&=E_{2, R}(t)-2\Delta t K_{1, I}E_{1, R}(t),\label{eq29c}\\
	E_{1, I}(t+\Delta t)&=E_{1, I}(t)+2\Delta t K_{1, I}E_{2, I}(t),\label{eq29d}
	\end{align}
\end{subequations}
and 
\begin{subequations}
	\begin{align}
	\Theta_D(\Delta t):\quad E_{2, R}(t+\Delta t)&=E_{2, R}(t),\label{eq30a}\\
	E_{1, I}(t+\Delta t)&=E_{1, I}(t),\label{eq30b}\\
	E_{1, R}(t+\Delta t)&=E_{1, R}(t)+2\Delta t K_{1, I}E_{2, R}(t),\label{eq30c}\\
	E_{2, I}(t+\Delta t)&=E_{2, I}(t)-2\Delta t K_{1, I}E_{1, I}(t).\label{eq30d}
	\end{align}
\end{subequations}
After getting the exact solutions individually, a simple first-order symplectic algorithm is just the composition of them
\begin{equation}
\Theta_1(\Delta t)=\Theta_R(\Delta t)\Theta_I(\Delta t)\Theta_A(\Delta t)\Theta_B(\Delta t)\Theta_C(\Delta t)\Theta_D(\Delta t),\label{eq31}
\end{equation}
where $\Theta_1(\Delta t)$ is the map of the numerical algorithm from $\mathbf{E}(t)$ to $\mathbf{E}(t+\Delta t)$ of order 1. 
The composition order of Eq.\eqref{eq31} can be either from left to right or vice versa, which means that the solving progress is divided into six steps and the solution of previous step is the initial value for the next step.

A second-order numerical symplectic scheme can be constructed by the composition of Eq.\eqref{eq31} and its adjoint that arranges the product of exact solutions in the inverse order, i.e.,
\begin{equation}
\begin{aligned}
\Theta_2(\Delta t)=&\Theta_R\left(\frac{\Delta t}{2}\right)\Theta_I\left(\frac{\Delta t}{2}\right)\Theta_A\left(\frac{\Delta t}{2}\right)\Theta_B\left(\frac{\Delta t}{2}\right)\Theta_C\left(\frac{\Delta t}{2}\right)\Theta_D(\Delta t)\\
&\Theta_C\left(\frac{\Delta t}{2}\right)\Theta_B\left(\frac{\Delta t}{2}\right)\Theta_A\left(\frac{\Delta t}{2}\right)\Theta_I\left(\frac{\Delta t}{2}\right)\Theta_R\left(\frac{\Delta t}{2}\right).
\label{eq32}
\end{aligned}
\end{equation}
Since all the subsystems have exact solutions, we are able to construct arbitrary high-order symplectic algorithms by compositing iteratively. One of the symplectic algorithm of order $2(l+1)$\cite{xiao_explicit_2015} is
\begin{equation}
\begin{aligned}
\Theta_{2(l+1)}(\Delta t)&=\Theta_{2l}(\alpha_l\Delta t)\Theta_{2l}(\beta_l\Delta t)\Theta_{2l}(\alpha_l\Delta t),\\
\alpha_l&=1/(2-2^{1/(2l+1)}),\\
\beta&=1-2\alpha_l,
\label{eq33}
\end{aligned}
\end{equation}
where $2(l+1)$ is the order of the symplectic algorithm.

\subsection{Numerical solution for nonlinear refractive index}
We discretize the four variables $E_{m, R},\ E_{m, I}\ (m=1, 2)$ on uniform grid within a two-dimensional computational domain $[x_L, x_R]\times[y_D, y_U]$ in a time $\tau$, with the spatial and temporal step size $\Delta x=\frac{x_R-x_L}{X},\ \Delta y=\frac{y_U-y_D}{Y}\ ,\Delta t=\frac{\tau}{T},\ \Delta t_f=\frac{\tau}{T_f}$, where $X, Y, T, T_f$ are positive integers. The $\Delta t$ and $\Delta t_f$ are the time step of CSEs and fluid equations respectively. Since the period of the EM wave is much shorter than the response time of the nonlinear medium generally, the $\Delta t_f$ depending on the property of the medium is much longer than $\Delta t$. In the following, we abbreviate the discrete $E_{m, R}^{t_n}(x_i, y_j)$ and $E_{m, I}^{t_n}(x_i, y_j)$ as $E_{m, R}^{n}(i, j)$ and $E_{m, I}^{n}(i, j)$. 

After discretization, the nonlinear refractive index in general media described by Eq.\eqref{eq14} can be obtained by substituting $E_{m, R},\ E_{m, I} (m=1, 2)$ with the discrete values $E_{m, R}^{n}(x_i, y_j)$ and $E_{m, I}^{n}(x_i, y_j)$.

However, if the nonlinear medium is a plasma, we need to solve Eq.\eqref{eq12a}-\eqref{eq12b}. The numerical difference scheme is to approximate the time derivatives by the forward Euler method, the spatial gradients by the centered finite difference and the divergences by the first-order upwind scheme.

For convenience, let $\phi=\bar{E}_1\bar{E}_2^{*}e^{-i\omega_3 t}+0.5(|\bar{E}_1^2+\bar{E}_2^2|)$ and the term on the righthand side of Eq.\eqref{eq12b} expressed with the discretized variable is 
\begin{equation}
\begin{aligned}
\phi _{R}^{n}=&\left( \bar{E}_{1,R}^{n}\bar{E}_{2,R}^{n}+\bar{E}_{1,I}^{n}\bar{E}_{1,I}^{n} \right) \cos \left( \bar{\omega}_3n\Delta t \right) +\left( \bar{E}_{1,I}^{n}\bar{E}_{2,R}^{n}-\bar{E}_{1,R}^{n}\bar{E}_{2,I}^{n} \right) \sin \left( \bar{\omega}_3n\Delta t \right),\\
& +0.5\left( |\bar{E}_{1}^{n}|^2+|\bar{E}_{2}^{n}|^2\right) \\
\phi _{I}^{n}=&\left( \bar{E}_{1,I}^{n}\bar{E}_{2,R}^{n}-\bar{E}_{1,R}^{n}\bar{E}_{2,I}^{n} \right) \cos \left( \bar{\omega}_3n\Delta t \right) -\left( \bar{E}_{1,R}^{n}\bar{E}_{2,R}^{n}+\bar{E}_{1,I}^{n}\bar{E}_{2,I}^{n} \right) \sin \left( \bar{\omega}_3n\Delta t \right) 
\end{aligned}
\end{equation}\label{eq34}
where $\phi_{R, I}^{n}$ denotes $\phi_{R,I}$ at $t=n\Delta t$ and $\omega_3$ is normalized by $\omega_1$. Likewise, we can choose the time $t_n$ to make $\phi_{R, I}^{N}$ synchronize with the fluid solver at $t=N\Delta t_f$  

The $\phi_{R, I}$ is the driving force of the fluid equations, with $\phi_{R, I}^N$ obtained from the CSEs solver, we solve the momentum equation Eq.\eqref{eq12b} first. Taking the curl of Eq.\eqref{eq12b} and letting $\nabla\cdot\mathbf{u_1}=\mathbf{U}$, we apply the finite difference to get the discretization of the equations. When $u_{0x}>0, u_{0y}>0$, it yields
\begin{equation}
	\begin{aligned}
	\frac{U_R^{N+1}(i,j)-U_R^{N}(i,j)}{\Delta t_f}&=-c_s^2\frac{n_{1, R}^{N}(i+1,j)+n_{1, R}^{N}(i-1,j)-2n_{1, R}^{N}(i,j)}{\Delta x^2}\\
	&-c_s^2\frac{n_{1, R}^{N}(i,j+1)+n_{1, R}^{N}(i,j-1)-2n_{1, R}^{N}(i,j)}{\Delta y^2}-2\nu_a U_R^{N}(i,j)\\
	&-u_{0x}\frac{U_R^{N}(i,j)-U_R^{N}(i-1,j)}{\Delta x}-u_{0y}\frac{U_R^{N}(i,j)-U_R^{N}(i,j-1)}{\Delta y}\\
	&-\frac{c_s^2}{4}\frac{\phi_R^{N}(i+1,j)+\phi_R^{N}(i-1,j)-2\phi_R^{N}(i,j)}{\Delta x^2}\\
	&-\frac{c_s^2}{4}\frac{\phi_R^{N}(i,j+1)+\phi_R^{N}(i,j-1)-2\phi_R^{N}(i,j)}{\Delta y^2},
	\end{aligned}\label{eq35}
\end{equation}
and 
\begin{equation}
	\begin{aligned}
	\frac{U_I^{N}(i,j)-U_I^{N}(i,j)}{\Delta t_f}&=-c_s^2\frac{n_{1, I}^{N}(i+1,j)+n_{1, I}^{N}(i-1,j)-2n_{1, I}^{N}(i,j)}{\Delta x^2}\\
	&-c_s^2\frac{n_{1, I}^{N}(i,j+1)+n_{1, I}^{N}(i,j-1)-2n_{1, I}^{N}(i,j)}{\Delta y^2}-2\nu_a U_I^{N}(i,j)\\
	&-u_{0x}\frac{U_I^{N}(i,j)-U_I^{N}(i-1,j)}{\Delta x}-u_{0y}\frac{U_I^{N}(i,j)-U_I^{N}(i,j-1)}{\Delta y}\\
	&-\frac{c_s^2}{4}\frac{\phi_I^{N}(i+1,j)+\phi_I^{N}(i-1,j)-2\phi_I^{N}(i,j)}{\Delta x^2}\\
	&-\frac{c_s^2}{4}\frac{\phi_I^{N}(i,j+1)+\phi_I^{N}(i,j-1)-2\phi_I^{N}(i,j)}{\Delta y^2}
	\end{aligned}\label{eq36}
\end{equation}
 With $U_{R, I}^N$, applying the finite difference sheme to Eq.\eqref{eq12a} allows us  to get the discretization as
 \begin{equation} 
 	\begin{aligned}
 	\frac{n_{1, R}^{N+1}(i,j)-n_{1,R}^N(i,j)}{\Delta t}=&-U_R^N(i,j)-u_{0x}\frac{n_{1, R}^{N}(i,j)-n_{1,R}^N(i-1,j)}{\Delta x}\\
 	&-u_{0,y}\frac{n_{1, R}^{N}(i,j)-n_{1,R}^N(i,j-1)}{\Delta y}
 	\end{aligned}\label{eq37}
 \end{equation}
 and
  \begin{equation} 
 	\begin{aligned}
 	\frac{n_{1, I}^{N+1}(i,j)-n_{1,I}^N(i,j)}{\Delta t}=&-U_I^N(i,j)-u_{0x}\frac{n_{1, I}^{N}(i,j)-n_{1,I}^N(i-1,j)}{\Delta x}\\
 	&-u_{0,y}\frac{n_{1, I}^{N}(i,j)-n_{1,I}^N(i,j-1)}{\Delta y}.
 	\end{aligned}\label{eq38}
 \end{equation}
 So the final discretization of the plasma fluid equations are Eq.\eqref{eq35} to Eq.\eqref{eq38} for the condition of $u_{0x}>0, u_{0y}>0$. Similarly, we can write the discretizations of other three cases when $u_{0x}>0, u_{0,y}<0$,  $u_{0x}<0, u_{0,y}<0$ and  $u_{0x}<0, u_{0,y}>0$ using the upwind scheme.
	\subsection{Initial and boundary conditions}
\begin{figure}[htbp]
	\centering
	\includegraphics[width=0.8\textwidth]{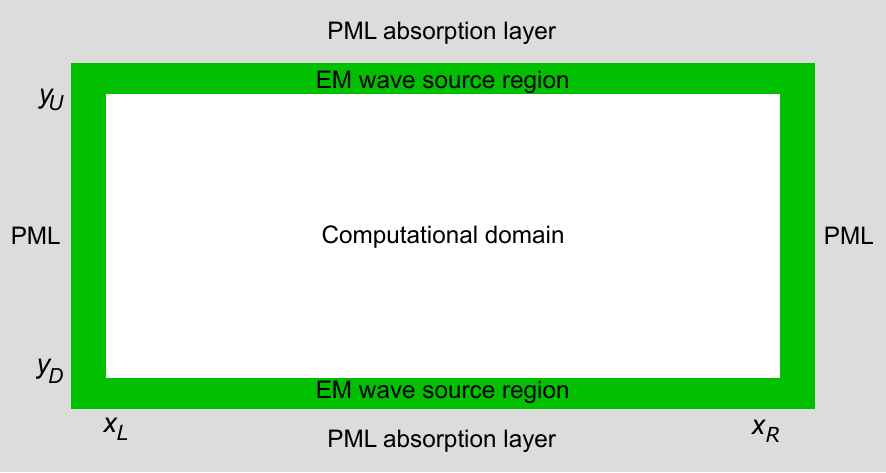}
	\caption{A schematic diagram of the computational domain, PML absorption layer and the region of wave sorce of electric fields. The relative position between the wave source and the absorption layer determines that the wave can only propagate inwards and is absorbed on the other side of the boundary. }
	\label{fig1}	
\end{figure}

The coupled Schrödinger equations describe the time evolution of a two-beam coupling system, once the initial and boundary conditions are given, we can solve the CSEs to predict the process at any future time.
As is shown in Fig.\ref{fig1}, the computational domain in which the CSEs are solved is in the center, the electromagnetic wave sources are loaded on the boundary of the computational domain (the green region of the figure), and the perfectly matched layer (PML)\cite{berenger_perfectly_1994} absorption layer is designed outside the region of wave sources. An exciting wave propagates to both sides but is absorbed in the PML region. The direction of the traveling wave is determined by the incident angle and injection position of the wave source, including top, bottom, left, and right.
In PML absorption region, regardless of the coupling terms, we only need to perform a complex coordinate stretching $x'\rightarrow x+i\beta(x)$ which evaluates the wave problem $e^{i(kx'-\omega t)}$ in complex space, and consequently, an exponentially decaying of the wave is introduced. To express the equations with respect to the real variable $x$, the complex part $i\beta(x)$ is transformed onto the complex material with an absorption coefficient. More details of the choice of absorption parameters have been discussed in Ref.\cite{berenger_perfectly_1994, zheng_perfectly_2007, nissen_optimized_2011}.
\section{Numerical results}
In this section, we present numerical examples to verify the validity of the numerical algorithm and show the performance of the developed code BEAM1D/2D. We demonstrate the Brillouin amplification and crossed-beam energy transfer using BEAM1D/2D and compare it with particle-in-cell (PIC) simulations.
\subsection{Brillouin amplification}
Brillouin amplification is a process that the energy of a long pump pulse is compressed by a counter-propagating short probe pulse. This process is under the strong-coupling regime when the plasma response is not an ion-acoustic mode of the plasma but a quisi-mode\cite{lancia_signatures_2016}. Here we consider the pump and probe light beams counter-propagate in a uniform and stationary plasma, which has the density $n_0=0.3n_c$ and the electron temperature $T_e=0.5\ \si{keV}$. The pump pulse has an intensity of $I_{pump}=10^{16}\ \si{W/cm^2}$ and a wavelength of $\lambda_1=1\ \si{\micro m}$, while the probe pulse of a temporal Gaussian profile has a peak intensity of $I_{probe}=10^{15}\ \si{W/cm^2}$ with a duration of 100 fs. The temperature ratio $ZT_e/T_i$ is set to 50, consequently, the Landau damping is negligible. The parameters chosen above are the same as the PIC simulation in Ref.\cite{andreev_short_2006}. As for BEAM1D, the spatial and temporal step size are set as $\Delta x=0.5c/\omega_1,\ \Delta t=0.1/\omega_1,\ \Delta t_f=40/\omega_1$ and the damping rate $\nu_a$ equals to zero.
\begin{figure}[htbp]
    \centering
    \subfigure[]{
        \includegraphics[width=0.8\textwidth]{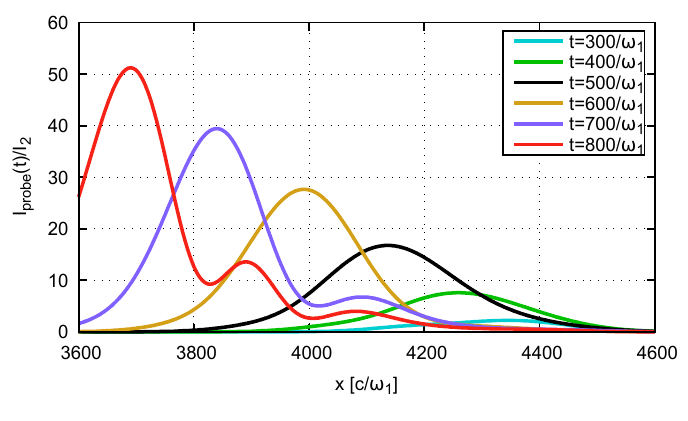}
        \label{fig2a}
    }
    \subfigure[]{
	\includegraphics[width=0.80\textwidth]{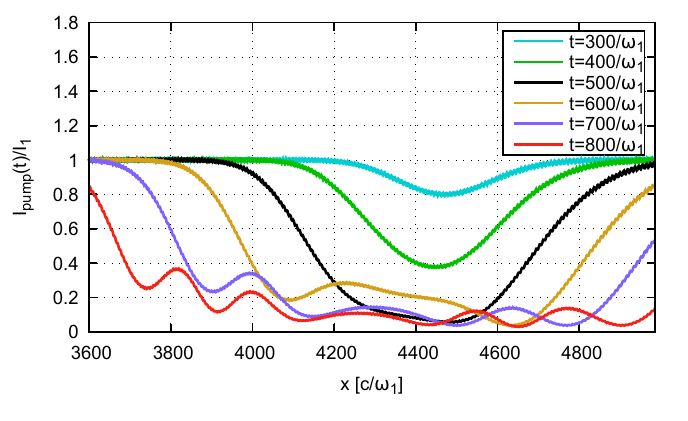}
        \label{fig2b}
    }
    \caption{BEAM1D results: intensity profiles of (a) the probe and (b) the pump pulses normalized by their respective original intensities $I_2=10^{15}\si{W/cm^2}$ and $I_1=10^{16}\si{W/cm^2}$ versus the spatial position $x$ normalized by $c/\omega_1$. In (a), at time $t=800/\omega_1$, the probe is amplified by a factor of about 50. The pulse width and the amplification factor at different times are quantitatively consistent with the PIC simulation results in FIG.2 of Ref.\cite{andreev_short_2006}. In (b), at the earlier time, the pump pulse is absorbed where it encounters the probe while at later times, the pump depletes distinctly.}
    \label{fig2}
\end{figure}

Figure \ref{fig2a} plots the intensity profiles of the seed pulse during Brillouin amplification, which characterizes by gradually higher and narrower intensity peaks. In \ref{fig2a}, at time $t=800/\omega_1$, the probe is amplified by a factor of about 50. The pulse width and the amplification factor at different times are quantitatively consistent with the PIC simulation results in FIG.2 of Ref.\cite{andreev_short_2006}. The depletion of the long pump pulse illustrated by Fig. \ref{fig2b} shows that at the early stage, the pump pulse is absorbed where it encounters the probe while at a later stage, the pump is depleted distinctly. In addition, the oscillatory behaviors of the pump and seed pulses demonstrate the self-similar ``$\pi-$pulse'' characteristic\cite{malkin_fast_1999}.
\subsection{Crossed-beam energy transfer}
In this subsection, we give the numerical results of CBET, including (i) two long pulses with different frequencies transferring energy in a stationary plasma and (ii) two long pulses of equal frequency interacting in a flowing plasma.

In the first case, we consider the energy transfer between two Gaussian laser beams ($w=15\si{\mu m}$) crossed at an angle of \ang{20}. The initial intensities are the same: $I_{pump}=I_{probe}=1.0\times 10^{15}\ \si{W/cm^{2}}$ and the pump beam has a wavelength $\lambda_1=1\si{\micro m}$ with the seed beam downshifted by an ion-acoustic wave frequency. The stationary helium plasma has an uniform density and temperature $n_0=0.04n_c,\ T_e=1.0\ \si{keV},\ T_i=0.333\ \si{\keV}$. The spatial and temporal step size of BEAM2D is $\Delta x=\Delta y=0.6283c/\omega_1, \Delta t=0.1/\omega_1, \Delta t_f=30/\omega_1$ and the damping rate is set as $\nu_a=0.05$ which is a little smaller than the linear Landau damping rate.
\begin{figure}[htbp]
	\centering
	\includegraphics[width=0.8\textwidth]{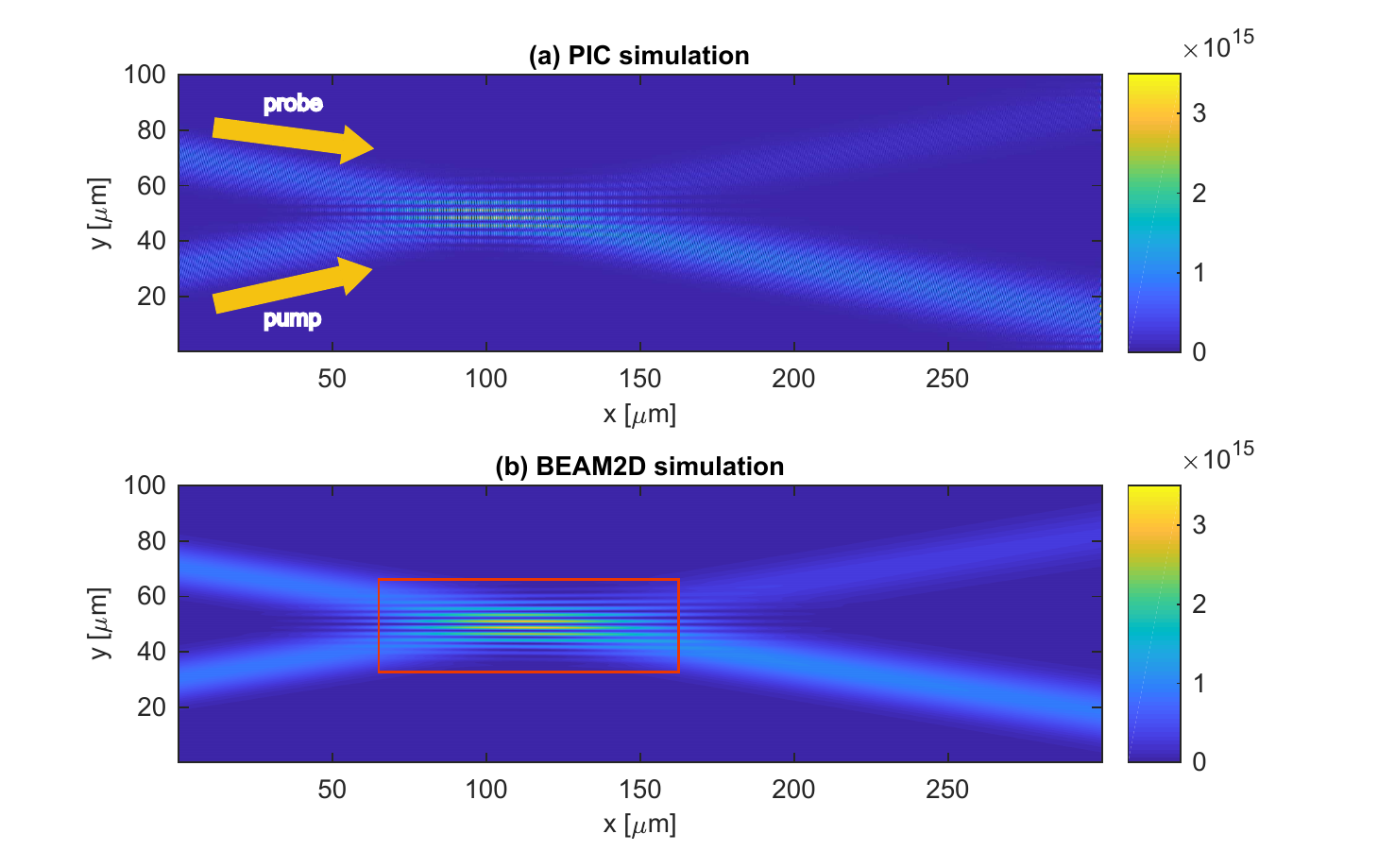}
	\caption{Intensity profiles of the superposition of the probe beam (incident from the upper left) and the pump beam (incident from the lower left) at $t=3900T (=13\si{ps}, \lambda_1=1\si{\mu m})$ obtained from (a) PIC simulation and (b) BEAM2D simulation. The rectangular box in Fig.(3b) marks the interaction region.}
	\label{fig3}
\end{figure}

Figure \ref{fig3} illustrates the intensity profiles of the superposition of the probe beam (incident from the upper left) and the pump beam (incident from the lower left) at $t=3900T (=13\si{ps})$ obtained from PIC code EPOCH\cite{arber_contemporary_2015} and BEAM2D code. Before the interaction region (marked by the rectangular box) where the interference pattern forms, the pump, and the probe beams have the same intensity, and after the interaction region, the intensity of the pump decreases while the seed intensity increases which illustrates an energy transfer. The transferred energy from the PIC code and BEAM2D are comparable.

\begin{figure}[htbp]
    \centering
    \subfigure[]{
        \includegraphics[width=0.8\textwidth]{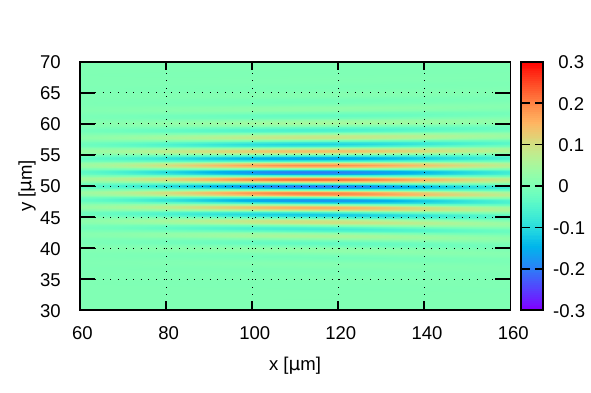}
        \label{fig4a}
    }
    \subfigure[]{
	\includegraphics[width=0.80\textwidth]{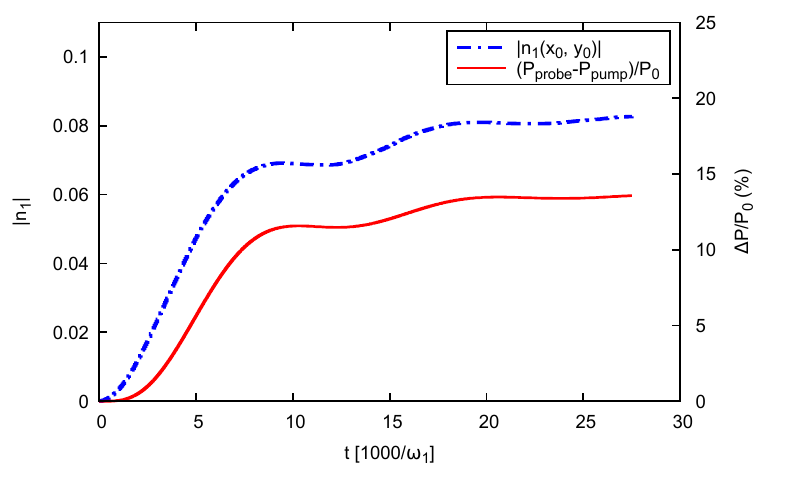}
        \label{fig4b}
    }
    \caption{Results of BEAM2D: (a) a snapshot of the ion density perturbation $\delta n$ in the interaction region (marked by the rectangular box in Fig.3(b) ) at $t=3900T (=13\si{ps})$. The excited ion-acoustic wave has a wavevector in the y-direction of $k_{iaw}=1.29k_0\approx 2k_0\si{sin10^{\circ}}$; (b) the temporal evolutions of the amplitude of the ion-acoustic wave (indicated by the blue dotted line) and the ratio of the transferred power to the initial power $\Delta P/P_0$ (indicated by the solid red line) during a period of 4700T ($\approx$ 16 ps).}
    \label{fig4}
\end{figure}
The snapshot of the ion density perturbation $\delta n$ located in the rectangular region of Fig.\ref{fig3} at $t=3900T (=13\si{ps})$ is plotted in Fig. \ref{fig4a}. The excited ion-acoustic wave with a wavevector in the y-direction of $k_{iaw}=1.29k_0\approx 2k_0\si{sin20^{\circ}}$, where $k_0$ is the wavevector of pump light in the plasma, satisfies the wavevector matching condition. In Fig.\ref{fig4b}, the temporal evolutions of the amplitude of the ion-acoustic wave (indicated by the blue dotted line) and the ratio of the transferred power to the initial power $\Delta P/P_0$ (indicated by the red solid line) during a period of 4700T ($\approx$ 16 ps) are shown.

Then we consider the second case where two long pulses with equal frequency counter-propagate in a flowing plasma. The two lasers have equal intensity $I=1\times10^{15}\ \si{W/cm^2}$ and the uniform hydrogen plasma density, normalized to the critical density, is $n_0=0.1n_c$. The temperatures of ion and electron are $T_i=1.5\ \si{keV}$ and $T_e=2.5\ \si{keV}$, respectively. Laser 1 propagates rightwards with a positive wavevector ($k_1=k_0$) while laser 2 propagates leftwards with a negative wavevector ($k_2=-k_0$). From the phase matching condition ($\omega_1-\omega_2=\omega_{iaw}=\mathbf{k_{iaw}\cdot u_0}\pm k_{iaw}c_s$), we know that a negative flowing velocity ($u_0<0$) is necessary for the energy transfer from laser 1 to laser 2 ($P_2-P_1>0$) case and $u_0=-c_s$ is at resonance. The deviation from resonance can be adjusted with different flowing velocities. According to the linear kinetic stationary model\cite{debayle_cross-beam_2018}, the practical ion-acoustic wave contains two damped modes and one driven mode. When near resonance, the driven mode reaches its maximum while the other modes almost vanish, however, when far from resonance, the competition between different modes will result in an oscillation of the energy transfer.

\begin{figure}[htbp]
    \centering
	\includegraphics[width=0.8\textwidth]{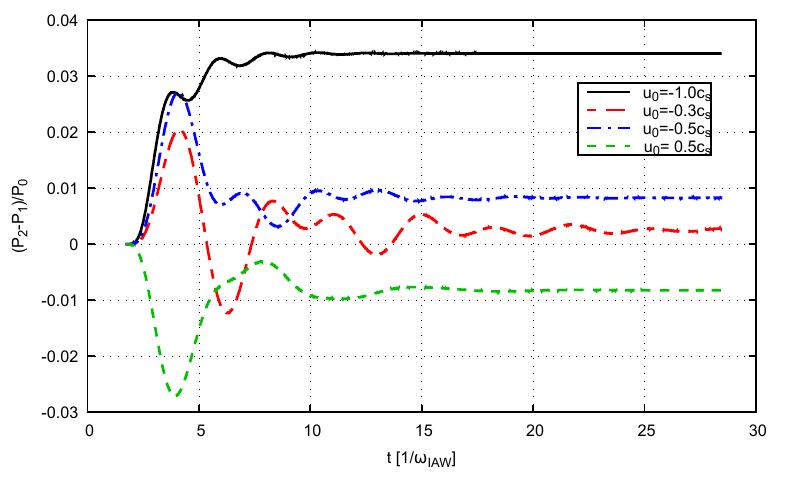}
    \caption{The transferred power (as a function of the initial power) aganist time (as a function of the reciprocal of the ion acoustic frequency $\omega_{iaw}$) under different plasma velocities.}
    \label{fig5}
\end{figure}

The growth and competition of different modes are demonstrated in \ref{fig5}, which illustrates the transferred power normalized by the initial power $P_0$ against time (as a function of the reciprocal of the ion acoustic frequency $\omega_{iaw}$) under different plasma velocities. When the resonance condition $u_0=-c_s$ (the black solid line) is satisfied, where the negative sign denotes the velocity in the negative direction of the x-axis, the energy transfers from laser 1 to laser 2 and the transferred energy exponentially increases initially then gradually saturates finally. When far from resonance, $u_0=-0.3c_s$, the direction of energy transfer changes alternatively, and the net transferred energy is still from laser 1 to laser 2, which is because of the competition between two waves with different frequencies ($\omega\propto k_{IAW}(u_0\pm c_s)\sim 0.7\omega_{IAW},\ 1.3\omega_{IAW}$) and hence the oscillation of the direction of energy transfer. The cases of opposite velocity indicate that the direction of net energy transferring varies with the direction of flowing velocity.

From the comparisons with kinetic simulations, we conclude that BEAM1D/2D can deal with the optical interactions in plasmas when the kinetic effect (ion-heating) is negligible, that is, when the laser intensities are moderate, or beams have a co-propagating configuration that will have minimal ion heating\cite{hansen_cross-beam_2022}.

\section{Conclusion}
Two laser beams can transfer energy in a nonlinear medium, and if one of them is a short pulse, the parametric amplification will occur. To describe this process, we derive a full-wave model, which can describe the coupling dynamics precisely by phase and amplitude. The model includes coupled Schrödinger equations and the equations of nonlinear complex refractive index. We find that the coupled Schrödinger equations form a Hamiltonian system. Consequently, we propose a fast semi-explicit symplectic scheme for long-time and large-scale simulations with a much lower computational cost compared to the PIC simulations. To demonstrate the validity of the numerical scheme, we benchmark the developed code BEAM1D/2D with PIC simulations.

Moreover, the numerical algorithm presented in this work could be extended to other nonlinear optical processes, such as stimulated Raman and Brillouin scattering, sum-frequency, and second harmonic generation. Because the phase-matching condition is generally necessary for these processes and hence the coupling coefficients are symmetric which makes it possible to find the Hamiltonian of the coupled-wave equations. From the perspective of the conservative system, the symplectic scheme ensures long-time numerical stability and therefore may have further extensive applications.

\section*{Acknowledgements}
This work is supported by the Strategic Priority Research Program of the Chinese Academy of Sciences, Grant No. XDA25010200. And we also thank the team of developers of the PIC code EPOCH.
\bibliography{JCPref}

\end{document}